\documentstyle[seceq,preprint,mbf]{ptptex}

\def\tr{\mathop{\rm tr}}
\def\slash#1{\ooalign{\hfil/\hfil\crcr$#1$}}

\def\XQ{{\mbf Q}}

\def\XD{{\mbf D}}
\def\XS{{\mbf S}}
\def\XK{{\mbf K}}

\def\dt{\!\cdot\!}
\def\nn{\nonumber\\}

\def\calA{{\cal A}}
\def\calD{{\cal D}}
\def\calN{{\cal N}}
\def\hatD{{\hat\calD}}

\def\hatR{{\hat R}}

\def\calF{{\cal F}}

\def\calR{{\cal R}}
\def\half{\hbox{\large ${1\over2}$}}
\def\myfrac#1#2{\hbox{\large ${#1\over#2}$}}
\def\T{{\rm T}}

\def\pr#1{{#1'}}
\def\subr#1{{#1_R\,}}

\def\4#1{{\underline{#1}}}
\def\m6#1{{\underline{#1}\,}}
\def\hA{{\cal A}}  \def\hF{{\cal F}}

\newdimen\Tdim
\def\ispan{{\setbox0=\hbox{i}%
\Tdim\ht0\advance\Tdim\dp0\rule[-\dp0]{0pt}{\Tdim}}}
\def\jspan{{\setbox0=\hbox{j}%
\Tdim\ht0\advance\Tdim\dp0\rule[-\dp0]{0pt}{\Tdim}}}
\def\Tspan#1{{\setbox0=\hbox{#1}%
\Tdim\ht0\advance\Tdim\dp0\advance\Tdim.55ex\rule[-\dp0]{0pt}{\Tdim}\box0}}
\def\abs#1{\left|#1\right|}
\def\calY{{\cal Y}}

\def\ul#1{{\underline{#1}}}
\def\XV{{\mbf V}}
\def\XL{{\mbf L}}
\def\XH{{\mbf H}}
\def\R{{\rm R}}
\preprintnumber[3cm]{
KUNS-1721\\ hep-th/0106051}

\title{
Off-Shell Formulation of Supergravity on Orbifold
}

\author{
Tomoyuki {\sc Fujita},\footnote{E-mail:
fujita@gauge.scphys.kyoto-u.ac.jp}
Taichiro {\sc Kugo}\footnote{E-mail:
kugo@gauge.scphys.kyoto-u.ac.jp}
and Keisuke {\sc Ohashi}\footnote{E-mail:
keisuke@gauge.scphys.kyoto-u.ac.jp}
}

\inst{
Department of Physics, Kyoto University, Kyoto 606-8502
}

\recdate{
\today
}

\abst{
An off-shell formulation is given for the 
``supersymmetry in singular 
spaces'' which has recently been developed in an on-shell formalism 
by Bergshoeff, Kallosh and Van Proeyen using supersymmetry singlet 
`coupling constant' field and 4-form multiplier field in 
five-dimensional space-time. We present this formulation for a general 
supergravity-Yang-Mills-hypermultiplet coupled system compactified on an
orbifold $S^1/Z_2$. Relations between the bulk cosmological constant and
brane tensions of the boundary planes are discussed. }

\begin{document}
\maketitle

\section{Introduction}

Concerning the supersymmetrization of the Randal-Sundrum 
scenario\cite{ref:RS2} on the orbifold $S^1/Z_2$,
there have appeared two distinct approaches; one is due to Altendorfer, 
Bagger and Nemeschansky\cite{ref:ABN,ref:GP} 
and another is due to Falkowski, Lalak and 
Pokorsky.\cite{ref:FLP} 
The difference resides in the point that the 
$U(1)_R$ gauge coupling constant $g$ and the gravitino `mass term' 
change their signs across the branes in the latter approach while they
do not in the former approach. (So the brane in the latter approach 
resembles to a thin limit of the domain wall.)
It is only in the latter approach that the supersymmetry requirement can 
give relations between the cosmological constant in the bulk space and 
the brane tensions of the two boundary planes. These relations are 
exactly the same relations which were required for the existence of the 
Randal-Sundrum's warp solution.\cite{ref:RS2} 
Moreover it is also the latter case that
is expected to appear from the heterotic M-theory on 
$S^1/Z_2$\cite{ref:HW} after the
reduction to five dimensions by compactifying on a 
Calabi-Yau 3-fold.\cite{ref:Lukas}

In a paper entitled ``Supersymmetry in Singular Spaces'', 
Bergshoeff, Kallosh and Van Proeyen\cite{ref:BKVP} 
(BKVP) have given an interesting formulation for realizing `dynamically' 
this situation of the coupling constant changing its sign across a 
brane. Namely, they considered the Maxwell/Einstein gauged supergravity 
system in five dimensions\cite{ref:GST} and {\em lifted} the gauge 
coupling constant $g$ of the $U(1)_R$ to a supersymmetry singlet {\it 
field} $G(x)$. Then, with introducing a 4-form gauge field 
$H_{\mu\nu\rho\sigma}$ also, they succeeded in constructing a 
supersymmetric action for the system on an $S^1/Z_2$ orbifold and 
realized the changing sign coupling `constant' $G(y)=g\epsilon(y)$ as 
the solution of the equation of motion.

However their construction is heuristic and is presented only in an
on-shell formulation for the pure Maxwell/Einstein gauged supergravity 
system. It is thus unclear how it becomes changed when the system is 
varied. The purpose of this paper is, therefore, to give an off-shell 
generalization for the BKVP formulation. In our formulation, the 
coupling field $G(x)$ appears as a ratio of the scalar component fields 
of two vector multiplets, and the 4-form gauge field 
$H_{\mu\nu\rho\sigma}$ is supplied essentially as a scalar component 
field of a linear multiplet. 

This paper is organized as follows. In \S2 we briefly explain the form 
of invariant off-shell action for the general 
supergravity-Yang-Mills-hypermultiplet system in 5D,\cite{ref:KO2} 
which we consider in this paper.
Next in \S3, we present a new form of linear multiplet in 
which the constrained vector and auxiliary scalar components are 
rewritten in terms of 3-form and 4-form gauge fields, respectively. This
was briefly pointed out in Ref.~\citen{ref:FO} but we present the 
details here for the first time. Using this new form of linear multiplet
as a supermultiplet containing BKVP's 4-form gauge field 
$H_{\mu\nu\rho\sigma}$,
we give an off-shell version of the 4-form gauge field in 5D bulk. In 
\S4 we discuss the compactification of the system on orbifold $S^1/Z_2$ 
and construct a brane action which again gives an off-shell 
generalization of BKVP's. In \S5, we discuss the relation between 
the cosmological constant and the brane tensions of the boundary planes, 
based on the obtained action. We show that various results obtained by 
previous authors are reproduced from our general results by reducing to 
simpler systems. Final section 6 is devoted to discussions. Some 
technical points are treated in Appendix concerning the parametrizations
of the target manifold $U(2,q)/U(2)\times U(q)$ of the hypermultiplet 
scalar fields and the non-linear Lagrangian. 

\section{Supergravity action in five-dimensional bulk}

The invariant action for a general system of Yang-Mills and 
hypermultiplet matters coupled to supergravity in off-shell formulation 
was first obtained in Ref.~\citen{ref:KO2}, which we refer to as I 
henceforth, based on the super Poincar\'e tensor calculus given in 
Ref.~\citen{ref:KO1}. However, the calculation was very tedious there 
since there was no conformal $S$-supersymmetry. Weyl multiplets in 5D 
conformal supergravity were constructed very recently by Bergshoeff et 
al\cite{ref:BCDWHP} and the full superconformal tensor calculus was 
presented by Fujita and Ohashi in Ref.~\citen{ref:FO}, which we refer to
as II, where it was explained how easily the result of I can be rederived 
based on the superconformal tensor calculus. We here, therefore, follow 
the technique developed in II.

We here consider the system of $n+1$ vector multiplets $\XV^I$ 
($I=0,1,2,\cdots,n$) of some gauge group $G$ and $r$ hypermultiplets 
$\XH^\alpha$ ($\alpha=1,2,\cdots,2r$) which give a certain 
representation of $G$ with representation matrix 
$(gt_I)^\alpha{}_\beta$. The field contents of the Weyl multiplet, 
vector multiplet and hypermultiplet are listed in Table.~\ref{table:1}.

\begin{table}[tb]
\caption{Field contents of the multiplets}
\label{table:1}
\begin{center}
\begin{tabular}{ccccc} \hline \hline
    field      & type   & restrictions & {\it SU}$(2)$ & Weyl-weight    \\ \hline 
\multicolumn{5}{c}{{\it Weyl multiplet}\phantom{MMM}} \\ \hline
\Tspan{$e_\mu{}^a$} &   boson    & f\"unfbein    & \bf{1}    &  $ -1$     \\  
$\psi^i_\mu$  &  fermion  & {\it SU}$(2)$-Majorana & \bf{2}
&$-\hspace{-1.5mm}\myfrac12$ \\  
\Tspan{$b_\mu$} & boson &  real & \bf{1} & 0 \\
\Tspan{$V^{ij}_\mu$}    &  boson    & $V_\mu^{ij}=V_\mu^{ji}=(V_{\mu ij})^*$ 
& \bf{3}&0\\ 
$v_{ab}$&boson& real, antisymmetric &\bf{1}&1 \\
$ \chi^i$  &  fermion  & {\it SU}$(2)$-Majorana & \bf{2}    &\hspace{-1.1mm}\myfrac32 \\  
$D$    &  boson    & real & \bf{1} & 2 \\ \hline
\multicolumn{5}{c}{{\it Vector multiplet $\XV^I$}\phantom{MMM}} \\ \hline
\Tspan{$W^I_\mu$}&  boson    & real gauge field   &  \bf{1}    &   0     \\
$M^I$& boson & real scalar& \bf{1} & 1 \\ 
$\Omega^{Ii}$      &  fermion  &{\it SU}$(2)$-Majorana  & \bf{2} &\myfrac32 \\  
$Y^I_{ij}$    &  boson    & $Y^{Iij}=Y^{Iji}=(Y^I_{ij})^*$  & \bf{3} & 2 \\ \hline
\multicolumn{5}{c}{{\it Hypermultiplet $\XH^\alpha$}\phantom{MMM}} \\ \hline
\Tspan{$\hA_i^\alpha$}     &  boson & 
$\hA^i_\alpha=\varepsilon^{ij}\hA_j^\beta\rho_{\beta\alpha}=-(\hA_i^\alpha)^*$ &\bf{2}& \myfrac32  \\  
$\zeta^\alpha$    &  fermion  & $\bar\zeta^\alpha\equiv(\zeta_\alpha)^\dagger\gamma_0 = \zeta^{\alpha\T}C$ 
& \bf{1}  & 2 \\ 
$\hF_i^\alpha$  &  boson    & 
$\hF^i_\alpha=-(\hF_i^\alpha)^*$  &  \bf{2}   & \myfrac52 \\ \hline
\end{tabular}
\end{center}
\end{table}

In the superconformal framework, the action obtained in I
results if we fix the extraneous gauge freedoms of dilatation $\XD$, 
conformal supersymmetry $\XS$ and special conformal-boost $\XK$ 
symmetries by the conditions\cite{ref:FO}
\begin{equation}
\XD:\ {\calN}=1,\qquad \XS:\ \Omega^{Ii}\calN_I=0,\qquad 
\XK:\  \hatD_a{\calN}=0,
\end{equation}
where $\calN(M)$ is the homogeneous cubic function of the
scalar component fields $M^I$ of the vector multiplets $\XV^I$ which 
uniquely characterizes the vector part action of the system. 
We use notations like $\calN_I\equiv\partial\calN/\partial M^I$, 
$\calN_{IJ}\equiv\partial\calN/\partial M^I\partial M^J$, etc, and $\hat\calD_\mu$ denotes full
superconformal covariant derivative. 
The $\XQ$ supersymmetry transformation which preserves these gauge 
conditions are given by the combination of 
$\XQ,\,\XS$ and $\XK$:
\begin{eqnarray}
\tilde \delta_Q(\varepsilon)&=&\delta_Q(\varepsilon)+\delta_S(\eta^i(\varepsilon))+\delta_K(\xi_K^a(\varepsilon)),\nn
\eta^i(\varepsilon)&=&-\myfrac{\calN_I}{12\calN}\gamma\dt \hat F^I(W)\varepsilon^i
+\myfrac{\calN_I}{3\calN}Y^{Ii}{}_j\varepsilon^j
+\myfrac{\calN_{IJ}}{3\calN}\Omega^{Ii}(2i\bar\varepsilon\Omega^J), 
\label{eq:Q}
\end{eqnarray}
omitting the expression of unimportant parameter $\xi_K^a(\varepsilon)$. 
The resultant $Q$ transformation laws of the Weyl multiplet, 
the vector multiplet and the hypermultiplet are completely the same as 
(I 6$\cdot$8), (I 6$\cdot$9) and (I 6$\cdot$10), respectively, given in 
I, provided that the following 
translation rules are used (the LHS is the present notations same as in II 
and the RHS is those in I)
\begin{eqnarray}
V_\mu^{ij}&\,\leftrightarrow\, &\tilde V_\mu^{ij}, \qquad 
v_{ab}\ \leftrightarrow\  \tilde v_{ab},\qquad 
\myfrac{\calN_I}{3\calN}Y^{Iij}\ \leftrightarrow\  -\tilde t^{ij},
\nn
\chi^i&\,\leftrightarrow\, &16\tilde \chi^i+3\gamma\dt \hat{\calR}^i(Q),
\qquad 
D\ \leftrightarrow\ 8\tilde C-\myfrac32\hat{\calR}(M)+2v^2,\nn
\Omega&\,\leftrightarrow\, &\lambda,\qquad 
\zeta_\alpha\ \leftrightarrow\ \xi_\alpha,\qquad 
Y^{Iij}\ \leftrightarrow\ \tilde Y^{Iij}-M^I\tilde t^{ij}.
\end{eqnarray}
We therefore omit those $\XQ$ transformation rules here, but 
cite the explicit form of the action for the present 
supergravity-Yang-Mills-hypermultiplet system:\cite{ref:KO2}
\begin{eqnarray}
&&\hspace{-5.5em}{\cal L}_0
={\cal L}_{\rm hyper}+{\cal L}_{\rm vector}+{\cal L}_{\rm C\hbox{-}S}+{\cal L}_{\rm aux}\ , \nn[1ex]
e^{-1}{\cal L}_{\rm hyper}&=&\nabla^a\calA_i^{\bar \alpha}\nabla_a\calA^i_\alpha 
-2i\bar\zeta^{\bar \alpha}(\slash{\nabla}+gM)\zeta_\alpha\nn
&&{}
+\calA^{\bar\alpha}_i(gM)^2{}_\alpha{}^\beta\calA_\beta^i
-4i\bar\psi^i_a\gamma^b\gamma^a\zeta_\alpha\nabla_b\calA^{\bar \alpha}_i 
-2i\bar\psi^{(i}_a\gamma^{abc}\psi_c^{j)}\calA^{\bar\alpha}_j\nabla_b\calA_{\alpha i}\nn
&&{}+\calA^{\bar\alpha}_i
\bigl(8ig\bar\Omega^i_{\alpha\beta}\zeta^\beta-4ig\bar\psi_a^i\gamma^aM_{\alpha\beta}\zeta^\beta 
\nn &&{}
\qquad \qquad +4ig\bar\psi_a^{(i}\gamma^a\Omega^{j)}_{\alpha\beta}\calA^\beta_j 
-2ig\bar\psi^{(i}_a\gamma^{ab}\psi^{j)}_bM_{\alpha\beta}\calA^\beta_j\bigr)
\nn
&&{}+\bar\psi_a\gamma_b\psi_c\bar\zeta^{\bar\alpha}\gamma^{abc}\zeta_\alpha 
-\myfrac12\bar\psi^a\gamma^{bc}\psi_a\bar\zeta^{\bar\alpha}\gamma_{bc}\zeta_\alpha\ ,\nn
e^{-1}{\cal L}_{\rm vector}&=&
  -\half R(\omega)-2i\bar\psi_\mu\gamma^{\mu\nu\rho}\nabla_\nu\psi_\rho 
  +(\bar\psi_a\psi_b)(\bar\psi_c\gamma^{abcd}\psi_d+\bar\psi^a\psi^b) \nn
&&{}-\calN_I \left(
   ig[\bar\Omega,\,\Omega]^I - \myfrac{i}4\bar\psi_c\gamma^{abcd}\psi_dF_{ab}(W)^I
\right)\nn
&&{}+ a_{IJ}
\left(\begin{array}{c}
-\myfrac14F(W)^I\dt F(W)^J+\myfrac12\nabla_aM^I\nabla^aM^J \\[.6ex]
+2i\bar\Omega^I\slash{\nabla}\Omega^J 
+i\bar\psi_a(\gamma\dt F(W)-2\slash{\nabla}M)^I\gamma^a\Omega^J \\[.5ex]
-2(\bar\Omega^{I}\gamma^a\gamma^{bc}\psi_a)(\bar\psi_b\gamma_c\Omega^{J})
+2(\bar\Omega^{I}\gamma^a\gamma^{b}\psi_a)(\bar\psi_b\Omega^{J})
\end{array}\right) \nn
&&{}-\calN_{IJK}\left(\begin{array}{c}
-i\bar\Omega^I\myfrac14\gamma\dt F(W)^J\Omega^K \\[.4ex]
{}+\myfrac23(\bar\Omega^I\gamma^{ab}\Omega^{J})(\bar\psi_a\gamma_b\Omega^{K}) 
+\myfrac23(\bar\psi^i\dt\gamma\Omega^{Ij})(\bar\Omega_{(i}^J\Omega_{j)}^K) 
\end{array}\right)\nn
&&
+\myfrac18\left(2\bar\psi_a\psi_b
+\bar\zeta^{\bar \alpha}\gamma_{ab}\zeta_\alpha 
+a_{IJ}\bar\Omega^I\gamma_{ab}\Omega^J\right)^2 \nn
&&{}+i\myfrac14\calN_IF(W)^I\left(2\bar\psi_a\psi_b
+\bar\zeta^{\bar \alpha}\gamma_{ab}\zeta_\alpha 
+a_{IJ}\bar\Omega^I\gamma_{ab}\Omega^J\right) \nn
&&{}+\bigl(\calA^{\bar\alpha i}\nabla_a\calA_\alpha^j
+ia_{IJ}\bar\Omega^{Ii}\gamma_a\Omega^{Jj}\bigr)^2 \nn
{\cal L}_{\rm C\hbox{-}S}&=&
\frac{1}8 c_{IJK}\epsilon^{\lambda\mu\nu\rho\sigma}W_{\lambda}^I
\bigl(F_{\mu\nu}^J(W)F_{\rho\sigma}^K(W) 
+\myfrac{1}2g[W_\mu,W_\nu]^JF_{\rho\sigma}^K(W) \nn
&&\hspace{8em}{}+\myfrac{1}{10}g^2[W_\mu,W_\nu]^J[W_\rho,W_\sigma]^K\bigr)\,. 
\label{eq:finalAction}
\end{eqnarray}
Here $\nabla_\mu$ denotes the derivative covariant only with respect to 
local-Lorentz and group transformations, and the metric $a_{IJ}$ in 
the vector multiplet kinetic term is 
\begin{equation}
a_{IJ}\equiv-{1\over2}{\partial^2\over\partial M^I\partial M^J}\ln\calN=-{1\over2\calN}
\bigl(\calN_{IJ}-{\calN_I\calN_J\over\calN}\bigr).
\label{eq:vecmetric}
\end{equation}
The barred group index $\bar\alpha$ of the hypermultiplet is defined to 
be $\calA^{\bar\alpha}\equiv\calA^{\beta}d_\beta{}^\alpha$ by using the 
hypermultiplet metric matrix $d_\alpha{}^\beta$ which is given in the 
standard form\cite{ref:dWLVP} as
\begin{equation}
d_\alpha{}^\beta=\pmatrix{{\bf{1}}_{2p}&  \cr
        &-{\bf{1}}_{2q} \cr}
\qquad \quad p+q=r.
\label{eq:hypmetric}
\end{equation}
Since this implies $\calA^{\bar\alpha}_i\calA_\alpha^i 
=-\sum_{\alpha=1}^{2p}|\calA^\alpha_i|^2+
\sum_{\alpha=2p+1}^{2(p+q)}|\calA^\alpha_i|^2$, 
the first $2p$ components of the hypermultiplet carry negative metric 
and are called compensator which will be eliminated eventually 
by the gauge-fixing of suitable gauge symmetries. We consider only 
$p=1$ and $p=2$ cases explicitly in this paper.

The last part of the lagrangian, ${\cal L}_{\rm aux}$, is 
the terms of the auxiliary fields which are written almost in the 
perfect square forms and vanish on shell, aside from the 
$Y^I_{ij}$-terms to which additional contributions will appear later:
\begin{eqnarray}
e^{-1}{\cal L}_{\rm aux}&=&
\pr D(\calA^2+2) -8i\bar{\pr \chi}{}^i\calA^{\bar\alpha}_i\zeta_\alpha 
+( Y^I_{ij}\hbox{-terms} )
\nn
&&{}+2(v-v_{\rm sol})^{ab}(v-v_{\rm sol})_{ab}
-(V_\mu-V_{\mu{\rm sol}})^{ij}(V^\mu-V^\mu_{\rm sol})_{ij} \nn
&&+\left(1-{A^2/\alpha^2}\right)
(\calF^{\bar\alpha}_i-\calF^{\bar\alpha}_{{\rm sol}\,i})
(\calF^i_\alpha-\calF^{\,i}_{{\rm sol}\,\alpha})\ , \nn
&&\hspace{-4.7em}(Y^I_{ij}\hbox{-terms})=-\myfrac12\calN_{IJ}Y^I_{ij}Y^{Jij}
+Y^I_{ij}\calY_I^{ij}\,,
\label{eq:Laux}
\end{eqnarray}
where
\begin{eqnarray}
\calY_I^{ij} &=&
2\calA^{(i}_{\alpha}(gt_I)^{\bar\alpha\beta}\calA_\beta^{j)}
+i\calN_{IJK}\bar\Omega^{Ji} \Omega^{Kj}, \nn
\pr D&\equiv&\myfrac18D+\myfrac3{16}\hatR(M)-\myfrac14v^2
-\myfrac{i}8\bar\psi\dt \gamma\chi 
+\myfrac{3i}8\bar\psi\dt \gamma\gamma \dt\hatR(Q)+i\bar\psi_a\gamma^{ab}{\mbf \Gamma}\psi_b, \nn
\pr \chi^i&\equiv&\myfrac1{16}\chi^i-\myfrac3{16}\gamma\dt\hatR^i(Q)
-\myfrac12\gamma^a{\mbf \Gamma}\psi_a^i, \nn 
{\mbf\Gamma}\varepsilon^i&\equiv& 
-3\eta^i(\varepsilon)
-\gamma^aV_a{}^i{}_j\varepsilon^j+\gamma\dt v\varepsilon^i\,. 
\label{eq:Psi}
\end{eqnarray}
with $\eta^i(\varepsilon)$ defined in Eq.~(\ref{eq:Q}). We have omitted the expressions for $V_{{\rm sol}\,a}^{ij}$, $v_{{\rm sol}\,ab}$ and $
\calF^{\,\alpha}_{{\rm sol}\,i}$ which are the same as given in I and 
we will not need below.

\section{The invariant action for the four-form gauge field in 5D
bulk}

A linear multiplet $\XL$ consists of 
an $SU(2)$ triplet boson $L^{ij}$, an $SU(2)$-Majorana spinor $\varphi^i$, 
a constrained vector $E^a$ and a real auxiliary scalar $N$.
An invariant action formula exists for a pair of an Abelian vector 
multiplet $\XV=(M,\,W_\mu,\,\Omega^i,\,Y^{ij})$ and a linear multiplet 
$\XL=(L^{ij},\,\varphi^i,\,E^a,\,N)$ which is neutral or charged under 
the Abelian group of the vector multiplet $\XV$:\cite{ref:FO,ref:KO1}
\begin{eqnarray}
e^{-1}{\cal L}_{\rm VL}\left(\XV, \XL\right) 
&=&Y^{ij} L_{ij}+2i\bar\Omega\varphi
 +2i\bar\psi^a_i\gamma_a\Omega_{j} L^{ij}\nn
&&-\myfrac12W_{a}\left(E^a-2i\bar\psi_b\gamma^{ba}\varphi +2i\bar\psi_b^{(i}\gamma 
 ^{abc}\psi_c^{j)}L_{ij}\right)\nn
&&+\myfrac12M \left(N-2i\bar\psi_b\gamma^{b}\varphi 
-2i\bar\psi_a^{(i}\gamma^{ab}\psi_b^{j)}L_{ij}\right). \label{eq:InvAction}
\end{eqnarray}
Now we consider the cases where the linear multiplet $\XL$ is 
neutral. It was observed in II that this lagrangian density can be 
consistently written as the following total derivative form of a 4-form
field $H_{\mu\nu\rho\sigma}$:
\footnote{Note, however, that we are {\em not} claiming that 
the action $\int d^5x\, {\cal L}_{\rm VL}(\XV, \XL)$ always vanishes. If the action 
$\int d^5x\, {\cal L}_{\rm VL}(\XV, \XL)$ is nonzero, then it merely implies that 
the 4-form field $H_{\mu\nu\rho\sigma}$ does not vanish at infinity so
that the surface term remains finite.}
\begin{equation}
2{\cal L}_{\rm VL}(\XV, \XL)=-\myfrac1{4!}\epsilon ^{\lambda\mu\nu\rho\sigma}\partial_\lambda(H_{\mu\nu\rho\sigma}
-4W_\mu E_{\nu\rho\sigma}), \label{eq:Nreplace}
\end{equation}
where $E_{\mu\nu\rho}$ is an unconstrained 3-form field with which the 
`divergenceless' constraint on $E^a$ is solved in the form\cite{ref:KO1}
\begin{eqnarray}
E^a&=&\myfrac1{4!}\epsilon ^{abcde}\hat F_{bcde}(E), 
\label{eq:EaCovEq}
\\
\hat F_{\mu\nu\rho\sigma}(E)&=&4\partial_{[\mu}E_{\nu\rho\sigma]}+
8i\bar\psi_{[\mu}\gamma_{\nu\rho\sigma]}\varphi 
+24i\bar\psi_{[\mu}^i\gamma_{\nu\rho}\psi_{\sigma]}^jL_{ij}.
\end{eqnarray}
The Eq.~(\ref{eq:Nreplace}) can be equivalently rewritten as
\begin{equation}
MN+2Y_{ij}L^{ij}+4i\bar\Omega\varphi =
-\myfrac1{5!}\epsilon ^{abcde}\hat F_{abcde}(H),\label{eq:NCovEq} 
\end{equation}
using the covariant field strength
$\hat F_{abcde}(H)$ of the 4-form field $H_{\mu\nu\rho\sigma}$:
\begin{eqnarray}
\hat F_{\lambda\mu\nu\rho\sigma}(H)&=&5\partial_{[\lambda}H_{\mu\nu\rho\sigma]}-10F_{[\lambda\mu}(W)E_{\nu\rho\sigma]}
-10i\bar\psi_{[\lambda}\gamma_{\mu\nu\rho\sigma]}M\varphi \nn
&&
+20i\bar\psi^i_{[\lambda}\gamma_{\mu\nu\rho\sigma]}\lambda^jL_{ij}
-40i\bar\psi^i_{[\lambda}\gamma_{\mu\nu\rho}\psi^j_{\sigma]}ML_{ij}.
\hspace{2em}
\end{eqnarray}
Note the parallelism between Eqs.~(\ref{eq:EaCovEq}) and (\ref{eq:NCovEq}).
Since the 4-form gauge field $H_{\mu\nu\rho\sigma}$ has 
$\left({4\atop4}\right)=1$ degree of freedom in 5D, it can be regarded as a 
replacement of the scalar component $N$ of the linear multiplet $\XL$, 
just as the 3-form gauge field $E_{\mu\nu\rho}$ (possessing 
$\left({4\atop3}\right)=4$ degrees of freedom in 5D) is the replacement 
of the constrained vector component $E^a$ of $\XL$. 
Therefore the linear multiplet 
$\XL=(L^{ij},\,\varphi^i,\,E^a,\,N)$ can now be expressed as
$\XL=(L^{ij},\,\varphi^i,\,E_{\mu\nu\rho},\,H_{\mu\nu\rho\sigma})$ by 
using the 3-form and 4-form gauge fields. 
The transformation 
$\delta\equiv\delta_Q(\varepsilon)+\delta_S(\eta)+\delta_E(\Lambda_{\mu\nu})+\delta_H(\Lambda_{\mu\nu\lambda})$ of 
these gauge fields are given by
\begin{eqnarray}
\delta E_{\mu\nu\lambda}&=&3\partial_{[\mu}\Lambda_{\nu\lambda]}
-2i\bar\varepsilon\gamma_{\mu\nu\lambda}\varphi 
-12i\bar\varepsilon^i\gamma_{[\mu\nu}\psi^j_{\lambda]}L_{ij},\nn
\delta H_{\mu\nu\rho\sigma}&=&4\partial_{[\mu}\Lambda_{\nu\rho\sigma]}
+6F_{[\mu\nu}(W)\Lambda_{\rho\sigma]}+2i\bar\varepsilon\gamma_{\mu\nu\rho\sigma}M\varphi \nn
&&{}-4i\bar\varepsilon^i\gamma_{\mu\nu\rho\sigma}\Omega^jL_{ij}
+16i\bar\varepsilon^i\gamma_{[\mu\nu\rho}\psi^j_{\sigma]}ML_{ij}+4(\delta_Q(\varepsilon)W_{[\mu})E_{\nu\rho\sigma]}.
\label{eq:EHtrf}
\hspace{2em}
\end{eqnarray}

It should, however, be kept in mind that this rewriting 
of the component field $N$ in terms of the 4-form gauge field 
$H_{\mu\nu\rho\sigma}$ is performed on the `background' not only of 
the Weyl multiplet but also of the vector multiplet $\XV$; 
that is, the $H_{\mu\nu\rho\sigma}$ component of $\XL$ depends on $\XV$, and we use the notation $H_{\mu\nu\rho\sigma}|_{\XV}$ and 
\begin{equation}
\XL|_{\XV} = (L^{ij},\,\varphi^i,\,
E_{\mu\nu\rho},\,H_{\mu\nu\rho\sigma}|_{\XV})
\end{equation}
to show explicitly the vector multiplet $\XV$ used in this rewriting. 

If we apply the invariant action 
formula ${\cal L}_{\rm VL}(\XV,\,\XL)$ in Eq.~(\ref{eq:InvAction}) 
using the same vector multiplet $\XV$ 
as that used in the rewriting $N \rightarrow H_{\mu\nu\rho\sigma}|_{\XV}$ of the 
linear multiplet $\XL$, the action takes the total derivative form
(\ref{eq:Nreplace}). However, we can use different vector multiplets,
$\XV_S$ and $\XV_R$, for the former and the latter,
respectively. Then the invariant action formula (\ref{eq:InvAction}) 
gives the desired 4-form field action:
\begin{eqnarray}
{\cal L}_{\rm 4\hbox{-}form}
&=&{\cal L}_{\rm VL}\left(\XV_S, \XL|_{\XV_R}\right) \nn
&=&e\Bigl((Y_S^{ij}-GY_R^{ij}) L_{ij}
+2i(\bar\Omega_S^i-G\bar\Omega_R^i) \varphi_i 
+2i\bar\psi^a_i\gamma_a(\Omega_{S\,j}-G\Omega_{R\,j}) L^{ij} \Bigr) \nn
&&{}-\frac1{4!}\epsilon^{\lambda\mu\nu\rho\sigma}\Bigl\{
\bigl(F_{\lambda\mu}(W_S)- G F_{\lambda\mu}(W_R)\bigr)E_{\nu\rho\sigma} 
+ \half G\,\partial_\lambda H_{\mu\nu\rho\sigma} \Bigr\}, 
\label{eq:4formAction}
\end{eqnarray}
where $G\equiv M_S/M_R$. 
This form is most easily obtained by rewriting as
\begin{equation}
{\cal L}_{\rm VL}\left(\XV_S, \XL|_{\XV_R}\right) =
{\cal L}_{\rm VL}\left(\XV_S-G\XV_R, \XL|_{\XV_R}\right) 
+G \,{\cal L}_{\rm VL}\left(\XV_R, \XL|_{\XV_R}\right). 
\end{equation}
Then, the first term can be calculated by 
the formula (\ref{eq:InvAction}) plaguing 
$\XV_S-G\XV_R=(0,\,W_{S\mu}-GW_{R\mu},\,\Omega_S^i-G\Omega_R^i,\,
Y_S^{ij}-GY_R^{ij})$ and using Eq.~(\ref{eq:EaCovEq}), and  
the second term gives $G$ times the total 
derivative form (\ref{eq:Nreplace}).

This Eq.~(\ref{eq:4formAction}) gives the desired 4-form field action,
which gives a off-shell generalization of the corresponding one by 
BKVP.\cite{ref:BKVP}
With this action, all the component fields 
of the linear multiplet $\XL|_{\XV_R}$ now play the role of 
Lagrange multipliers; in particular, 
the variation of the 4-form field $H_{\mu\nu\rho\sigma}$ 
constrains the field ratio $G(x)\equiv M_S(x)/M_R(x)$ to be a constant 
which plays the role of a coupling constant $g_R$:
\begin{equation}
\partial_\mu G(x)=0 \  \rightarrow\  G(x) = g_R\ ({\rm constant}). 
\end{equation}
The other equations given by the variation of other components read
\begin{equation}
Y_S^{ij}= G Y_R^{ij}, \qquad 
\Omega_{S\,j}=G\Omega_{R\,j}, \qquad 
F_{\lambda\mu}(W_S)= G F_{\lambda\mu}(W_R).
\label{eq:multiplierEq}
\end{equation}

It would be appropriate to explain here how the gauged $U(1)_R$ 
supergravity is constructed with these two vector multiplets $\XV_S$ and
$\XV_R$ in the present formulation. 
We identify the vector multiplet 
$\XV_R$ with the $U(1)_R$ gauge multiplet which is generally given by a 
linear combination of the (Abelian) vector multiplets $\XV^I$:
\begin{equation}
\XV_R\equiv V_I\XV^I, \quad {\rm i.e.}\quad 
(M_R, \ W_{R\mu},\ \cdots\ )=
(V_IM^I, \ V_IW^I_\mu,\ \cdots\ ).
\end{equation}
We use the index $I$ to denote all the vector multiplets other than 
$\XV_S$. 
In the paper I in which no $\XV_S$ appeared, the $U(1)_R$ gauge 
multiplet $\XV_R\equiv V_I\XV^I$ was made couple {\it only} to the 
hypermultiplet compensator in the form
\begin{equation}
\calD_\mu\calA^{a}_i =
\partial_\mu\calA^{a}_i -V_{\mu ij}\calA^{a j}
-g_RW_{R\mu}(i\vec q\dt\vec\sigma)^{a }{}_{b}\calA^{b}_i
\label{eq:CovHyp0}
\end{equation}
where the indices $a ,\, b$ denote 1 and 2, the first two values 
of $\alpha=1,2,\cdots$ corresponding to the first compensator and $\vec q$ is 
any constant isovector of unit length, $|\vec q\,{}^2|=1$. 
Then, after the $SU(2)$ gauge of the index $i$ is fixed by the condition
$\calA^a_i\propto\delta^a_i$, the kinetic term $-|\calD_\mu\calA^a_i|^2$ of 
the compensator $\calA^a_i$ gives 
a square term in $V^{\rm N}_\mu{}^{ij}=V_\mu^{ij}-g_RW_{R\mu}(i\vec q\dt\vec\sigma)^{ij}$ so that the auxiliary $SU(2)$ gauge field $V_\mu^{ij}$ contained 
in the covariant derivative $\calD_\mu$ in all the other places 
become replaced by 
$V^{\rm N}_\mu{}^{ij}+g_RW_{R\mu}(i\vec q\dt\vec\sigma)^{ij}$, thus yielding a 
universal coupling of $W_{R\mu}$ to the $U(1)_R$ subgroup of $SU(2)$ with 
index $i$. (See I for more details.) 

Here, on the other hand, $\XV_S$ eventually becomes the $U(1)_R$ gauge
multiplet $\XV_R$ times the `coupling constant' $G=g_R$ by the equations of 
motion (\ref{eq:multiplierEq}), and thus the $\XV_S$ may be called 
`pre-$U(1)_R$ gauge multiplet'.  In the present formulation, therefore, 
it is this pre-$U(1)_R$ gauge multiplet $\XV_S$ that we make couple to 
the hypermultiplet compensator at the beginning: namely, we have 
\begin{equation}
\calD_\mu\calA^{a}_i =
\partial_\mu\calA^{a}_i -V_{\mu ij}\calA^{a j}
-W_{S\mu}(i\vec q\dt\vec\sigma)^{a}{}_{b}\calA^{b}_i
\label{eq:CovHyp}
\end{equation}
which reduces to the previous Eq.~(\ref{eq:CovHyp0}) after the equations
of motion (\ref{eq:multiplierEq}) are used. 
We assume that the pre-$U(1)_R$ gauge 
multiplet $\XV_S$ does not have its own kinetic term; that is, $M_S$ is 
not contained in $\calN(M)$. The kinetic term of the $U(1)_R$ gauge 
multiplet, of course, exists when $\det \calN_{IJ}\not=0$ which we assume 
throughout this paper.

\section{Compactifying on orbifold and the brane action}

We now compactify the fifth direction $y\equiv x^4$ on the orbifold 
$S^1/Z_2$, the two fixed planes of which are placed at $y=0$ and 
$y=\pi R\equiv\tilde y$. 
We must first know the properties of the fields 
under the $Z_2$ parity transformation, $y \rightarrow-y$. The parity quantum 
number $\Pi$ is defined by 
\begin{equation}
\Phi(-y) = \Pi(\Phi) \, \Phi(y)
\end{equation}
for boson fields $\Phi$, and, as discussed by BKVP,\cite{ref:BKVP} by
\begin{equation}
\psi^i(-y) = \Pi(\psi)\, \gamma_5 M^i{}_j \psi^j(y) 
\qquad 
(\bar\psi^i(-y)=\Pi(\psi) M^i{}_j \bar\psi^j(y)\gamma_5), 
\end{equation}
for $SU(2)$ spinor fermions $\psi^i$ ($i=1,2$). 
Consistency with the reality $\bar\psi^i=(\psi_i)^\dagger\gamma_0=\psi^{i\,\T}C$ requires 
\begin{equation}
(M^i{}_j)^* = \epsilon_{ik}M^k{}_l\epsilon^{lj}, \qquad 
\hbox{or}\quad    
M^* = -\sigma_2M\sigma_2
\end{equation}
and we can take 
\begin{equation}
M^i{}_j = 
(\sigma_3)^i{}_j =\delta^i_j(-1)^{j+1}
\end{equation}
without loss of generality.  For fermion components $\zeta^\alpha$ 
$(\alpha=1,2,\cdots,2r)$ of hypermultiplets, the parity is similarly defined by
\begin{equation}
\zeta^\alpha(-y) = \Pi(\zeta)\, \gamma_5 M^\alpha{}_\beta\zeta^\beta(y) 
\end{equation}
and so the reality condition demands 
$(M^\alpha{}_\beta)^* = \rho_{\alpha\gamma}M^\gamma{}_\delta\rho^{\delta\beta}$. Thus we can take 
$M^\alpha{}_\beta= \sigma_3 \otimes {\bf 1}_r$ in the standard representation in which 
$\rho_{\alpha\beta}= \epsilon \otimes {\bf 1}_r$. 

The parity is determined by demanding the invariance of the action and 
the consistency of the both sides of the supersymmetry transformation rules.
We find, for the Weyl multiplet fields and $Q$- and 
$S$-transformation parameters $\varepsilon$ and $\eta$,
\begin{eqnarray}
&&\Pi(e_{\4\mu}^{\4 a})= \Pi(e_y^4)= +1, \quad 
\Pi(e_{\4\mu}^4)=\Pi(e_y^{\4a})=-1, \quad  \nn
&&\Pi(\psi_{\4\mu}) = \Pi(\varepsilon)= +1, \quad 
\Pi(\psi_y) = \Pi(\eta)= -1, \quad \nn
&&\Pi(b_{\4\mu}) = \Pi(V_{\4\mu}^3)= \Pi(V_y^{1,2}) = \Pi(v^{4{\4 a}}) 
= \Pi(\chi)=\Pi(D)= +1, \nn
&&\Pi(b_y) = \Pi(V_y^3)= \Pi(V_{\4\mu}^{1,2})= \Pi(v^{{\4 a}{\4 b}})= -1,
\end{eqnarray}
where the underlined indices 
$\ul\mu$ and $\ul a$ denote 
the four-dimensional parts of 
the five-dimensional curved index $\mu$ and flat index $a$. The fifth 
direction of $\mu$ and $a$ are denoted by $y$ and $4$, respectively.
The (real) `isovector' components $\vec t=(t^{1,2,3})$ is 
generally defined for any symmetric $SU(2)$ tensor $t^{ij}$ (satisfying 
hermiticity $t^{ij}=(t_{ij})^*$) as 
\begin{equation}
t^i{}_j (= t^{ik}\varepsilon_{kj}) \equiv i\vec t \cdot \vec\sigma^i{}_j.
\end{equation}
For vector multiplet $\XV=(M,\,W_\mu,\,\Omega^{i},\,Y^{ij})$, we have
\begin{eqnarray}
&&\Pi(M) = \Pi(W_y) = \Pi(Y^{1,2})= \Pi_{\XV}, \nn
&&\Pi(\Omega) = \Pi(W_{\4\mu}) = \Pi(Y^{3})= -\Pi_{\XV}.
\end{eqnarray}
We define the parity $\Pi_{\XV}$ of vector multiplet $\XV$ to be 
the parity $\Pi(M)$ of the first component scalar $M$. 
Normally the parity $\Pi_{\XV}$ of vector multiplets must be $+1$ in 
five dimensions since they appear in the action via the homogeneous 
cubic function $\calN(M)$ which should have even parity $+1$. However, 
if a certain subset of the vector multiplets appear in $\calN(M)$ 
only in the terms quadratic in them, then their parity assignment 
has a choice of $\pm1$. 

For linear multiplet 
$\XL=(L^{ij},\,\varphi^i,\,E^a,\,N)$, we find
\begin{eqnarray}
&&\Pi(L^{1,2}) = \Pi(\varphi) = \Pi(N) = \Pi(E^4) 
= \Pi(E_{\4\mu\4\nu\4\rho})= \Pi_{\XL}, \nn
&&\Pi(L^{3}) = \Pi(E^{\4 a}) = \Pi(E_{y\4\mu\4\nu}) = -\Pi_{\XL}\,.
\end{eqnarray}
The parity of the 4-form field $H_{\mu\nu\rho\sigma}|_{\XV}$ also depends on 
the parity $\Pi_{\XV}$ of the vector multiplet $\XV$ used in the 
rewriting $N \rightarrow H_{\mu\nu\rho\sigma}$:
\begin{equation}
\Pi(H_{\4\mu\4\nu\4\rho\4\sigma})= -\Pi_{\XV}\Pi_{\XL}, \qquad 
\Pi(H_{y\4\mu\4\nu\4\rho})= \Pi_{\XV}\Pi_{\XL}.
\end{equation}

The hypermultiplet $\XH^\alpha=(\calA^\alpha_i,\,\zeta^\alpha,\, \calF^\alpha_i)$ 
$(\alpha= 1,2,\cdots,2r)$ splits into $r$ pairs 
$(\XH^{2\hat\alpha-1}, \XH^{2\hat\alpha})$ $(\hat\alpha=1,2,\cdots,r)$ 
in the standard representation in which 
$\rho_{\alpha\beta}= \epsilon \otimes {\bf 1}_r$ and then the 
reality condition of the scalar components $\calA^\alpha{}_i$, 
$\epsilon^{ij}\calA^\beta{}_j\rho_{\beta\alpha}=(\calA^\alpha{}_i)^*$, implies that the 
$2\times2$ matrix $(\calA^{2\hat\alpha-1}{}_i, \calA^{2\hat\alpha}{}_i)$ with 
$i=1,2$ for each fixed $\hat\alpha$ (which can be identified with a 
quaternion\footnote{Quaternion ${\mbf q}=
\calA^0_{\hat\alpha}-{\mbf i}\calA^1_{\hat\alpha}
-{\mbf j}\calA^2_{\hat\alpha}-{\mbf k}\calA^3_{\hat\alpha}$}
) has the form
\begin{equation}
\pmatrix{\calA^{2\hat\alpha-1}{}_1 & \calA^{2\hat\alpha-1}{}_2 \cr
\calA^{2\hat\alpha}{}_1 & \calA^{2\hat\alpha}{}_2 \cr}
=\calA^0_{\hat\alpha}\,{\bf1}_2 + \sum_{k=1}^3 i\calA^k_{\hat\alpha} \,\sigma_k =
\pmatrix{\calA^0_{\hat\alpha}+i\calA^3_{\hat\alpha} & 
i\calA^1_{\hat\alpha}+\calA^2_{\hat\alpha} \cr
 i\calA^1_{\hat\alpha}-\calA^2_{\hat\alpha} & 
\calA^0_{\hat\alpha}-i\calA^3_{\hat\alpha} \cr}.
\label{eq:quaternion}
\end{equation}
with $\calA^{0,1,2,3}_{\hat\alpha}$ being real. The components 
$\calF^{0,1,2,3}_{\hat\alpha}$ are also defined similarly. 
Then, if the component $\calA^0_{\hat\alpha}$ 
has a parity $\Pi_{\hat\alpha}$, then we have 
\begin{equation}
\Pi(\calA^{0,3}_{\hat\alpha})=\Pi(\calF^{1,2}_{\hat\alpha}) 
= \Pi_{\hat\alpha}, \qquad 
\Pi(\calA^{1,2}_{\hat\alpha})=\Pi(\zeta_{2\hat\alpha-1,2\hat\alpha}) 
= \Pi(\calF^{0,3}_{\hat\alpha}) = -\Pi_{\hat\alpha}.
\label{eq:hypP}
\end{equation}

In the present gauged $U(1)_R$ supergravity, the 
vector multiplet $\XV_S$ couples to the hypermultiplet compensator 
in the form (\ref{eq:CovHyp}). 
In order for the two terms $-V_{\mu ij}\calA^{a j}$ and 
$-W_{S\mu}(i\vec q\dt\vec\sigma)^{a }{}_{b }\calA^{b }_i$ 
in the RHS of Eq.~(\ref{eq:CovHyp}) to have the same $Z_2$ parity
property, $\vec q\dt\vec\sigma$ must commute or anti-commute with $\sigma_3$, 
and $\Pi(W_{S\4\mu}) =+1$ when $[\vec q\dt\vec\sigma,\,\sigma_3]=0$ and $\Pi(W_{S\4
\mu})=-1$ when $\{\vec q\dt\vec\sigma,\,\sigma_3\}=0$. This implies that the 
parity $\Pi_S$ of the vector multiplet $\XV_S$ (i.e., $\Pi(M_S)$) should 
be
\begin{equation}
\Pi_S = \cases{
-1  & when $\vec q\dt\vec\sigma=\sigma_3$ \cr
+1  & when $\vec q\dt\vec\sigma=\sigma_1\cos\theta+\sigma_2\sin\theta$ \cr}
\end{equation}
In order for the 4-form field action 
${\cal L}_{\rm VL}\left(\XV_S, \XL|_{\XV_R}\right)$ 
in Eq.~(\ref{eq:4formAction}) to be invariant under $Z_2$-parity, 
we must have $\Pi_S\Pi_{\XL}=+1$. 

As discussed by BKVP, we want to have the pullback of the component 
$H_{\4\mu\4\nu\4\rho\4\sigma}$ on the brane nonvanishing so that 
$\Pi(H_{\4\mu\4\nu\4\rho\4\sigma})=+1$. This implies $\Pi_R\Pi_{\XL}=-1$ and hence
$\Pi_R=-\Pi_S$ and $\Pi(G)=-1$. 

We allow the cubic term in $M_R$ to exist in $\calN(M)$, thus allowing 
the Chern-Simons term of the gauge field $W_{R\mu}$ in the five dimensional 
bulk, we assign $\Pi_R=+1$, then we have $\Pi_S=-1$ and $\Pi_{\XL}=-1$. 
Note that, since $\Pi_S=-1$, the vector multiplet $\XV_S$ or its first 
component $M_S$ can appear at most only in quadratic form in $\calN(M)$. 
For simplicity, we assume that $\XV_S$ does not appear in $\calN(M)$ 
at all, as announced before.

With these parity quantum numbers kept in mind, 
we consider the transformation rules (\ref{eq:EHtrf}) of 
$H_{\mu\nu\rho\sigma}$ and 
\begin{equation}
\delta M =2i\bar\varepsilon\Omega, \qquad \quad 
\delta L^{ij}=2i\bar\varepsilon^{(i}\varphi ^{j)},
\end{equation}
for the first scalar components $M$ and $L^{ij}$ of vector and linear 
multiplets. Keeping all the even parity fields nonvanishing on the brane
in the RHSs of these equations, we find the following transformation 
rules on the brane:
\begin{eqnarray}
\delta H_{\4\mu\4\nu\4\rho\4\sigma}&=&4\partial_{[\4\mu}\Lambda_{\4\nu\4\rho\4\sigma]}
+2i\bar\varepsilon\gamma_{\4\mu\4\nu\4\rho\4\sigma}\varphi\, M_R \nn
&&{}-4\bar\varepsilon\gamma_{\4\mu\4\nu\4\rho\4\sigma}\gamma_5\Omega_R\,L^3
+16\bar\varepsilon\gamma_5\gamma_{[\4\mu\4\nu\4\rho}\psi_{\4\sigma]}M_R L^3,\nn
\delta M_\R&=&2i\bar\varepsilon\Omega_{R},\qquad \delta L^3=\bar\varepsilon\gamma_5\varphi.
\end{eqnarray}
Here we have used an identity 
$\bar\varepsilon\gamma_5\varphi =-2\bar\varepsilon^{(1}\varphi^{2)}$ which holds on the brane 
for $SU(2)$ Majorana 
spinors $\varepsilon^i$ and $\varphi ^i$ with $\Pi(\varepsilon)=+1$ and $\Pi(\varphi)=-1$.
Now it is clear that the following brane action is superconformal 
invariant:
\begin{eqnarray}
S_{\rm brane}=
\int d^5x\left(\Lambda_1\delta(y)+\Lambda_2\delta(y-\tilde y)\right)\left[
\myfrac1{4!}\epsilon ^{\mu\nu\rho\sigma y}H_{\mu\nu\rho\sigma}+2e_{(4)}M_\R L^3 \right]
\label{eq:braneAction}
\end{eqnarray}
where $e_{(4)}\equiv e/e_y^4$ is the determinant of the 
four dimensional vierbein on the brane. 


\section{Relation between bulk cosmological constant and brane tensions}

Now the action of our total system is given by the bulk action 
${\cal L}_0$ [Eq.~(\ref{eq:finalAction})] plus the 4-form field action 
${\cal L}_{\rm 4\hbox{-}form}$ [Eq.~(\ref{eq:4formAction})] plus 
the brane action ${\cal L}_{\rm brane}$ [Eq.~(\ref{eq:braneAction})]. 
The latter two parts ${\cal L}_{\rm 4\hbox{-}form} + {\cal L}_{\rm brane}$
give an off-shell generalization of the BKVP action\cite{ref:BKVP} 
for realizing the odd parity coupling `constant' (field) $G(x)$. 

Indeed, variation of the components $\varphi^i$ and $E_{\mu\nu\rho}$ of 
the linear multiplet still gives the same equations as in 
Eq.~(\ref{eq:multiplierEq}): $\Omega_{S\,j}=G\Omega_{R\,j}$ and $F_{\lambda\mu}(W_S)= 
G F_{\lambda\mu}(W_R)$. The latter implies that $W^\mu_S= G W_R^\mu$ up to a 
gauge transformation if $G$ is a constant. This equation clearly shows that 
the field $G(x)$ plays the role of $U(1)_R$ gauge coupling constant.
On the other hand, the equation obtained by 
variation of the 4-form field $H_{\mu\nu\rho\sigma}$ is now changed 
in the presence of branes into
\begin{equation}
\partial_\mu G=
2\delta_\mu^y\left(\Lambda_1\delta(y)+\Lambda_2\delta(y-\tilde y)\right).
\end{equation}
The integrability condition of this equation on the orbifold $S^1/Z_2$ 
demands the condition 
\begin{equation}
\Lambda_1=-\Lambda_2\equiv\subr g,
\end{equation}
and then the solution of $G(y)$ is given by
\begin{equation}
G=\subr g\epsilon(y)
\end{equation}
with a periodic sign function $\epsilon(y)$ (with period $2\tilde y$) 
which is defined as
\begin{equation}
\epsilon(y)\equiv 
\cases{
+1 &for $0<y<\tilde y$ \cr
-1 &for $-\tilde y<y<0$ \cr
0 &for $y=0,\,\tilde y$ \cr}.
\end{equation}
That is, the coupling `constant' $G$ changes its sign across the branes 
and is really constant away from the branes.  

Let us now discuss the relation between the cosmological constant in 5D 
bulk and the brane tensions of the boundary planes. 
First note that the auxiliary fields $Y^{ij}$ are contained in the 
action in the form [See (\ref{eq:Laux}) and (\ref{eq:4formAction})]
\begin{equation}
-\myfrac12\calN_{IJ}Y^I_{ij}Y^{Jij}+Y^I_{ij}\calY^{ij}_I 
+Y^S_{ij}\calA^i_{a }(2gt_S)^{a b }\calA^j_{b }
+(Y^S-GY^R)_{ij}L^{ij}
\label{eq:5.1}
\end{equation}
where $a,\,b=1,2$ are the indices of the first compensator $\calA^a_i$ 
only to which the vector multiplet $\XV_S$ couples, 
$(gt_S)^a{}_b=(i\sigma_3)^a{}_b$\footnote{%
We adopt the convention that the usual expression for the Pauli matrix 
$\vec\sigma$ applies to the matrix with index position $\vec\sigma^i{}_j$. 
The indices $i$ and $j$ are raised or lowered by using $\epsilon$ tensor,
so that, for instance,  $\vec\sigma^{ij}=\epsilon^{jk}\vec\sigma^i{}_k=
-\vec\sigma^i{}_k\epsilon^{kj}$ denotes 
$-\vec\sigma\epsilon$ as a matrix. }
and the last term came from 
${\cal L}_{\rm 4\hbox{-}form}$. The vector multiplet $\XV_S$ is special since 
it is not contained in $\calN(M)$ and so has no kinetic term. Then the 
auxiliary field component $Y_S$ appears in the action only linearly and 
plays a role of Lagrange multiplier yielding a constraint equation:
\begin{eqnarray}
&&L^{ij} 
= -2
\calA^{i}_{a }(i\sigma_3)^{a b }\calA^{j}_{b }
\equiv L^{ij}_{\rm sol} , \quad \hbox{or equivalently} \nn
&&L^1=L^2=0 \quad {\rm and}\quad 
L^3= 
-2i\calA^{1}_{a }(i\sigma_3)^{a b }\calA^{2}_{b }
\equiv L^3_{\rm sol}.
\label{eq:L3sol}
\end{eqnarray}
Then the Eq.~(\ref{eq:5.1}) is rewritten into
\begin{eqnarray}
&&-\myfrac12\calN_{IJ}(Y-Y'_{\rm sol})^I_{ij}(Y-Y'_{\rm sol})^{Jij}
+\myfrac12\calN_{IJ}Y^{\prime\,I}_{{\rm sol}\,ij}Y^{\prime\,Jij}_{\rm sol}\nn
&&\qquad Y_{\rm sol}^{\prime\,Iij}\equiv 
(\calN^{-1})^{IJ}(\calY_J^{ij}-GV_JL_{\rm sol}^{ij})
\label{eq:Ysquare}
\end{eqnarray}

After the auxiliary fields $Y_S$, $\varphi^i$, $E_{\mu\nu\rho}$,
$H_{\mu\nu\rho\sigma}$ and $Y^I$s are eliminated, the brane action now takes 
the form
\begin{equation}
S_{\rm brane}=
-g_R\int d^5x\left(\delta(y)-\delta(y-\tilde y)\right)\left[
3e_{(4)}W L^3_{\rm sol} \right],
\end{equation}
where $W\equiv-(3/2)M_R=-(3/2)V_IM^I$ is  the `superpotential' introduced 
later in Eq.~(\ref{eq:superpot}) below. This shows that brane tensions 
of the planes at $y=0$ and $y=\tilde y$ are given by 
$\mp 3g_RWL^3_{\rm sol}$, respectively.

Next we turn to compute the scalar potential in 5D bulk. For that purpose, 
it is better to discuss separately two cases of one and two compensators. 
As noted before,  the hypermultiplet 
scalars $\calA^\alpha_i$ ($\alpha=1,2,\cdots,2r$) (or fermions $\zeta^\alpha$) 
have the metric matrix $d_\alpha{}^\beta$ given in (\ref{eq:hypmetric}) 
in their kinetic term, and so
the first $2p$ components $\calA^\alpha_i$ with $\alpha=1,2,\cdots,2p$, 
corresponding to $p$ quaternions, have negative metric and are called 
compensators, while the rest components $\calA^{\ul\alpha}_i$ which we 
denote with underlined index $\ul\alpha$ have positive metric and represent 
usual matter fields. We consider the simplest two cases of one ($p=1$) and 
two ($p=2$) compensators, separately. 

\subsection{$p=1$ case}

First is the most common $p=1$ case, containing a single 
compensating hypermultiplet $\XH^a$ with $a=1,2$. 
Independently of $p$, we always fix the $SU(2)$ gauge by imposing the 
condition 
\begin{equation}
\calA^a_i(x) = a(x) \delta^a_i , \qquad a(x): \hbox{real positive}
\label{eq:su2GF}
\end{equation}
on the first compensator scalars $\calA^{a=1,2}_i$, and so the 
$L^3_{\rm sol}$ in Eq.~(\ref{eq:L3sol}) takes the form, 
\begin{equation}
L^3_{\rm sol}= 
-2i\calA^{1}_{a }(i\sigma_3)^{a b }\calA^{2}_{b }
=-2(\sigma_3)^1{}_1a^2(x)=-2a^2(x).
\end{equation}
In the present case of 
$p=1$, the equation of motion $\calA^2+2=0$\footnote{%
Note that the four compensating bosons $\calA^a_i$ 
($a=1,2$, $i=1,2$) carrying negative metric in this $p=1$ case are 
eliminated by 
the three gauge-fixing conditions of $SU(2)$, (\ref{eq:su2GF}), and 
this equation of motion $\calA^2=-2\calN$ with dilatation gauge 
condition $\calN=1$. The target
manifold spanned by the hypermultiplet scalars constrained by 
these conditions becomes $USp(2,2q)/USp(2)\times USp(2q)$.\cite{ref:BSVP}}
for $\calA^2 \equiv\calA^\alpha_id_\alpha{}^\beta\calA_\beta^i= 
-\abs{\calA^a_i}^2+\abs{\calA^{\ul\alpha}_i}^2$ determines the 
magnitude $a(x)$ to be
\begin{equation}
a(x)=\sqrt{1+\myfrac12|\calA_i^{\underline{\alpha}}|^2}.
\label{eq:aI}
\end{equation}
The scalar potential in the bulk is given by\cite{ref:KO2} 
\begin{eqnarray}
V&=&\left.-\myfrac12\calN_{IJ}\pr Y^I_{{\rm sol}ij}\pr Y^{Jij}_{\rm sol}
\right|_{\rm bosonic}
-\calA^{\bar\alpha}_i(gM)^2{}_\alpha{}^\beta\calA_\beta^i \nn
&=&(a^{IJ}-M^IM^J)P_I^{ij}(P^{ij}_J)^*
-|Q^a_i|^2+ |Q^{\ul\alpha}_i|^2 
\label{eq:scalarpot}
\end{eqnarray}
where use has been made of $(\calN^{-1})^{IJ}=-(1/2)(a^{IJ}-M^IM^J)$ 
with $a^{IJ}$ denoting the inverse of $a_{IJ}$ in 
Eq.~(\ref{eq:vecmetric}), and taking Eq.~(\ref{eq:Ysquare}) into 
account, 
\begin{eqnarray}
P_I^{ij} &\equiv& -\half GV_IL_{\rm sol}^{ij}+\half\calY^{ij}_I|_{\rm bosonic}
=GV_Ia^2(i\sigma_3)^{ij}
+\calA^{(i}_{\ul\alpha}(gt_I)^{\ul\alpha\ul\beta}\calA_{\ul\beta}^{j)}
, \nn
Q^a_i&\equiv&\delta_G(M)\calA^a_i= M_S(gt_S)^a{}_b\calA^b_i 
= GM_R(i\sigma_3)^a{}_b\calA^b_i
= GV_IM^Ia(i\sigma_3)^a{}_i, \nn
Q^{\ul\alpha}_i&\equiv&\delta_G(M)\calA^{\ul\alpha}_i
=M^I(gt_I)^{\ul\alpha}{}_{\ul\beta}\calA^{\ul\beta}_i. 
\end{eqnarray}
If the hypermultiplet fields $\calA^{\ul\alpha}_i$ other than the compensator 
$\calA^a_i$ are assumed not to be charged, i.e., $gt_I=0$, then 
the potential reduces to a simple form
\begin{eqnarray}
V&=&\myfrac32 g_R^2\left\{
3\abs{a}^4g^{xy}{\partial W\over\partial\varphi^x}{\partial W\over\partial\varphi^y}
-(\abs{a}^4+3\abs{a}^2)W^2\right\}.
\label{eq:scalarpotI}
\end{eqnarray}
Here $\varphi^x$ ($x=1,\cdots,n$) are independent $n$ scalar fields with
which $n+1$ vector multiplet scalars $M^I$, constrained by $\XD$ gauge 
condition $\calN(M)=1$, are parametrized.\cite{ref:GST} 
We have used the relations 
$a^{IJ}=g^{xy}h^I_xh^J_y+h^Ih^J$ and $h^I=-\sqrt{2/3}M^I$ given in I, 
Eqs.~(I 7$\cdot$3) and (I 7$\cdot$1), and the definitions\cite{ref:BKVP}
\begin{equation}
W\equiv\sqrt{\myfrac23}V_Ih^I = -\myfrac23V_IM^I\,, \qquad 
{\partial W\over\partial\varphi^x}
= -\myfrac23V_Ih^I_x 
= -\myfrac23V_IM^I_{,x}\,.
\label{eq:superpot}
\end{equation}
If the system contains no physical vector multiplets (i.e., $n=0$) 
so that there 
appear no scalars $\varphi^x$ and only the graviphoton with $I=0$ exists 
as a vector field, then, the `superpotential' $W$ reduces to a 
constant and the bulk scalar potential $V$ to 
$V=-(3/2)g_R^2(\abs{a}^4+3\abs{a}^2)W^2$. This should be compared with the 
brane tensions $\pm6g_RaW$ of the planes at $y=0$ and $\tilde y$. 
If the system has no matter-hypermultiplets either, i.e., $q=0$, then,
$a$ becomes 1 and the bulk scalar potential further reduces to 
$V=-6g_R^2W^2$ and brane tensions to $\pm6g_RW$, yielding 
exactly the same relations required by the 
Randal and Sundrum.\cite{ref:RS2}

\subsection{$p=2$ case}

Next we turn to $p=2$ case, for which the manifold spanned by the 
hypermultiplet scalar fields becomes $SU(2,q)/SU(2)\times SU(q)\times 
U(1)$ and, when $q=1$, this just corresponds to the manifold of the 
universal hypermultiplet\cite{ref:CFG} appearing in the reduction of the
heterotic M-theory on $S^1/Z_2$ to five dimensions.\cite{ref:Lukas} 
In this $p=2$ case, 
as explained in I in detail, we need to introduce another special vector
multiplet $\XV_T$ which possesses no kinetic term either and couples to 
the hypermultiplets via charge $T=\sigma_3\otimes{\bf1}_{2+q}$, that 
is, the hypermultiplet $\XH^\alpha$ with odd (even) $\alpha$ carries 
$+1$ ($-1$) charge. In this case the auxiliary field component 
$Y_T^{ij}$ again appears in the action as a multiplier field and gives 
three constraints
\begin{equation}
\calA_{ai}(\sigma_3)^a{}_b\calA^b_j
+\calA_{a'i}(\sigma_3)^{a'}{}_{b'}\calA^{b'}_j
=\calA_{\ul\alpha i}(\sigma_3)^{\ul\alpha}{}_{\ul\beta}\calA^{\ul\beta}_j
\label{eq:YTconstr}
\end{equation}
where the primed indices $a',\,b'=3,\,4$ are used to denote those of the
second compensator $\calA^{a'}_i$ and the indices $\alpha,\,\beta$ of 
$(\sigma_3)^\alpha{}_\beta$ should generally be understood to be 1 and 
2 when they are odd and even, respectively. The equation of motion 
$\calA^2 \equiv\calA^\alpha_id_\alpha{}^\beta\calA_\beta^i =-2$ gives 
\begin{equation}
-\abs{\calA^a_i}^2
-|\calA^{a'}_i|^2
+\abs{\calA^{\ul\alpha}_i}^2 = -2.
\label{eq:EQmotion}
\end{equation}
If we impose the $SU(2)$ gauge-fixing condition Eq.~(\ref{eq:su2GF}) and 
$\calA^{a'=3}{}_{i=2}=$ real as the $U(1)_T$ gauge-fixing, the solution of 
these constraints  (\ref{eq:YTconstr}) and (\ref{eq:EQmotion}) is
shown in the Appendix to be given in terms of two $q$-component 
complex vectors $\phi_1$ and $\phi_2$ as
\begin{eqnarray}
&&\calA^{a=1}{}_{i=1}
= \sqrt{ 1-\abs{\phi_2}^2\over2(1-\abs{\phi_1}^2-\abs{\phi_2}^2
+\abs{\phi_1}^2\abs{\phi_2}^2-
|\phi_1^\dagger\cdot\phi_2|^2) } \equiv a, \label{eq:aII} \\
&&
\calA^{a'=3}{}_{i=1}=
a{\phi_2^\dagger\cdot\phi_1\over1-\abs{\phi_2}^2},
\qquad 
\calA^{a'=3}{}_{i=2} 
= \sqrt{1\over2(1-\abs{\phi_2}^2)}\equiv b, \nn
&&\calA^{\ul\alpha={\rm odd}}{}_{i=1}= 
a\left(\phi_1+{\phi_2^\dagger\cdot\phi_1\over1-\abs{\phi_2}^2}\phi_2\right), 
\qquad 
\calA^{\ul\alpha={\rm odd}}{}_{i=2}= b\phi_2,
\label{eq:solA}
\end{eqnarray}
where $\abs{\phi}^2\equiv\phi^\dagger\cdot\phi$ and we have shown only 
the components $\calA^\alpha_i$ with odd $\alpha=2\hat\alpha-1$, whose 
complex conjugates essentially give the even 
$\alpha=2\hat\alpha$ components, 
$\calA^{2\hat\alpha}{}_i = (\calA^{2\hat\alpha-1}{}_j)^*\epsilon_{ji}$,
because of the quaternionic nature (\ref{eq:quaternion}) of 
the hypermultiplet scalars $\calA^\alpha_i$. 

The scalar potential in the bulk is now given by 
\begin{eqnarray}
V&=&
(a^{IJ}-M^IM^J)P_I^{ij}(P^{ij}_J)^*
-|Q^a_i|^2 
-|Q^{a'}_i|^2+ |Q^{\ul\alpha}_i|^2, 
\label{eq:scalarpotII}
\end{eqnarray}
where $P_I^{ij}$ is the same as before while $Q^\alpha_i$ are now
\begin{eqnarray}
Q^a_i
&\equiv&\delta_G(M)\calA^a_i=
(M_S+M_T)(i\sigma_3)^a{}_b\calA^b_i 
= (GV_IM^I+M_T)a(i\sigma_3)^a{}_i, \nn
Q^{a'}_i&\equiv&\delta_G(M)\calA^{a'}_i=M_T(i\sigma_3)^{a'}{}_{b'}\calA^{b'}_i, 
\nn
Q^{\ul\alpha}_i&\equiv&\delta_G(M)\calA^{\ul\alpha}_i
=\bigl(M_T(i\sigma_3)^{\ul\alpha}{}_{\ul\beta}
+M^I(gt_I)^{\ul\alpha}{}_{\ul\beta}\bigr)\calA^{\ul\beta}_i. 
\end{eqnarray}
If we assume again that 
the hypermultiplet matter fields $\calA^{\ul\alpha}_i$ other than the 
compensators $\calA^a_i$ and $\calA^{a'}_i$ are not charged, 
i.e., $gt_I=0$, 
then the potential takes a simpler form. The contribution of the 
three $\abs{Q^\alpha_i}^2$ terms is evaluated as
\begin{eqnarray}
&&\hspace{-3em}-|Q^a_i|^2 
-|Q^{a'}_i|^2+ |Q^{\ul\alpha}_i|^2  \nn
&=&
-2(M_S+M_T)^2\abs{a}^2 
-M_T^2|\calA^{a'}_i|^2+ M_T^2|\calA^{\ul\alpha}_i|^2 \nn
&=&
-2M_S^2\abs{a}^2 
-4M_SM_T\abs{a}^2 
+M_T^2(-|\calA^{a}_i|^2-|\calA^{a'}_i|^2+|\calA^{\ul\alpha}_i|^2) \nn
&=&
-2M_S^2\abs{a}^2 
-4M_SM_T\abs{a}^2 
-2M_T^2
\label{eq:5.23}
\end{eqnarray}
where we have inserted the Eq.~(\ref{eq:EQmotion}) in the last step.
Since $\XV_T$ has no its own kinetic term and the scalar component 
$M_T$ appears only here in the action, so that $M_T$ can be eliminated 
by using its equation of motion $M_T=-M_S\abs{a}^2$. Then this 
contribution (\ref{eq:5.23}) reduces to
\begin{equation}
-2M_S^2\abs{a}^2 +2M_S^2\abs{a}^4.
\end{equation}
The first term of this is the same contribution as the previous $p=1$ 
case, which together with the first $|P^{ij}_I|^2$ term gives the 
same expression as the previous potential (\ref{eq:scalarpotI}) 
(although $a$ here is given by Eq.~(\ref{eq:aII}) and different from the 
previous one (\ref{eq:aI}) ).
Thus the second term is 
the additional new contribution in this $p=2$ case, 
\begin{equation}
2M_S^2\abs{a}^4 = 2g_R^2(V_IM^I)^2\abs{a}^4 
= \myfrac92g_R^2W^2\abs{a}^4. 
\end{equation}
Adding this to the previous potential (\ref{eq:scalarpotI}), we find 
the scalar potential in $p=2$ case as 
\begin{eqnarray}
V&=&\myfrac32 g_R^2\left\{
3\abs{a}^4g^{xy}{\partial W\over\partial\varphi^x}{\partial W\over\partial\varphi^y}
-(3\abs{a}^2-2\abs{a}^4)W^2\right\}.
\label{eq:scalarpotIII}
\end{eqnarray}

Again consider the special case that the system contains no 
physical vector multiplets and so no scalars $\varphi^x$, then, $W$ is 
constant. Moreover, consider the case $q=1$; that is, the system contains 
only a single physical hypermultiplet and corresponds to the universal 
hypermultiplet. Then, $\phi_1$ and $\phi_2$ are single 
component complex fields and can be rewritten in terms of more popular 
variables, a complex $\xi$ and real $V$ and $\sigma$:\cite{ref:BPSV}
\begin{equation}
\phi_1={2\xi\over1+S}, \qquad 
\phi_2={1-S\over1+S}, \qquad  S\equiv V+\bar\xi\xi +i\sigma. 
\label{eq:param}
\end{equation}
Then, $a$ in Eq.~(\ref{eq:aII}) reduces to
\begin{equation}
a=\sqrt{{1\over2}+{\abs{\xi}^2\over2V}}.
\label{eq:q1a}
\end{equation}

We can easily see that the result (\ref{eq:scalarpotIII}) with this $a$ 
in (\ref{eq:q1a}) is easily seen to agree with the result by Falkowski, 
Lalak and Pokorski.\cite{ref:FLP} 
Actually the present result (\ref{eq:scalarpotIII}) 
reproduces only a part of their result, the part proportional to their $
\beta$ parameter, but the part proportional to their $\alpha$ parameter is 
missing. The reason is clear. Up to here we have tacitly assumed that 
our pre-$U(1)_R$ gauge multiplet $\XV_S$ couples only to the first 
compensator. Namely the $U(1)_S$ charge $gt_S$ to which $\XV_S$ 
couples has been chosen to be 
\begin{equation}
gt_S = i\pmatrix{
1 & 0 & 0 \cr
0 & 0 & 0 \cr
0 & 0 & 0 \cr}\qquad {\rm on}\quad 
\calA_{\rm odd} \equiv\Bigl( \calA^{2\hat\alpha-1}{}_i \Bigr)
\equiv\pmatrix{ 
\calA^{a=1}{}_{i}\cr
\calA^{a'=3}{}_{i} \cr
\calA^{\ul\alpha=5}{}_{i} \cr} .
\label{eq:former}
\end{equation}
However, the isometry group of the hypermultiplet manifold is 
$U(2,q{=}1)$, which is given by the $3\times3$ matrices $U_{2,1}$ acting
on this $3\times2$ matrix $\calA_{\rm odd}$ from left (and its complex 
conjugate $(U_{2,1})^*$ on the even row elements $\calA_{\rm 
even}\equiv(\calA^{2\hat\alpha}{}_i)$). When the system is 
compactified on $S^1/Z_2$, the isometry group is reduced to the subgroup
$U(1)\times U(1,q{=}1)$ since it should commute with the $Z_2$ parity 
transformation under which $(\calA^{a=1}{}_{1}, 
\calA^{a'=3}{}_{1},\, \calA^{\ul\alpha=5}{}_{1})$ get signs $(+1, -1, 
-1)$. (This parity assignment corresponds to $\Pi(\phi_1)=-1$ and 
$\Pi(\phi_2)=+1$ (or, $\Pi(\xi)=-1$ and $\Pi(S)=+1$) and is consistent 
with Eq.~(\ref{eq:hypP}) and expressions of $\calA^{2\hat\alpha-i}{}_i$ 
in Eqs.~(\ref{eq:aII}) and (\ref{eq:solA}).) The $U(1)_S$ charge 
$gt_S$ to which $\XV_S$ couples can actually be any $U(1)$ generator of 
this isometry group $U(1)\times U(1,q{=}1)$. 
The $U(1)_S$ generator which was chosen by Falkowski 
et al is given, in our terminology, by 
\begin{equation}
gt_S = i\alpha\pmatrix{
0 & 0 & 0 \cr
0 & -1 & -1 \cr
0 & 1 & 1 \cr} + 
i\beta\pmatrix{
1 & 0 & 0 \cr
0 & 0 & 0 \cr
0 & 0 & 0 \cr}.
\label{eq:latter}
\end{equation} 
The first generator with coefficient $\alpha$ corresponds to the 
isometry in $U(1,q{=}1)$ shifting the field $\sigma={\rm Im}\,S$ in 
Eq.~(\ref{eq:param}) and the second generator with coefficient $\beta$ 
is the original $U(1)$ charge in Eq.~(\ref{eq:former}). Also in this 
case, one can compute the bulk cosmological constant and brane tensions 
in the same way as above. It is convenient to use the $3\times2$ matrix 
$\Phi\equiv\calA_{\rm odd}$ consisting of the odd row elements 
(\ref{eq:former}) and notation defined in Appendix, with which we find 
formulas ($W_{,x}\equiv\partial W/\partial\varphi^x$) 
\begin{eqnarray}
&&L_{\rm sol}{}^i{}_j 
=4 \bigl(\Phi^\dagger\eta(gt_S)\Phi 
-\half {\bf1}_2\tr(\Phi^\dagger\eta gt_S\Phi)\bigr)^i{}_j \nn
&&(a^{IJ}-M^IM^J)P_I^{ij}(P^{ij}_J)^*
= \myfrac3{16}g_R^2\bigl(
3g^{xy}W_{,x}W_{,y}-W^2\bigr)\tr(L_{\rm sol}^\dagger L_{\rm sol}) \nn
&&-|Q^a_i|^2 
-|Q^{a'}_i|^2+ |Q^{\ul\alpha}_i|^2
=-2M_S^2\bigl\{
\tr(\Phi^\dagger\eta(gt_S)^2\Phi)+
\bigl[\tr(\Phi^\dagger\eta gt_S\Phi)\bigr]^2\bigr\} \nn
&&\hspace*{13em}
+2\bigl(M_T+iM_S\tr(\Phi^\dagger\eta gt_S\Phi)\bigr)^2
\end{eqnarray}
The last perfect square term vanishes if we use $M_T$ equation of motion. 
Inserting the expression (\ref{eq:latter}) for the $U(1)_S$ generator 
$gt_S$ in this case and 
expressions (\ref{eq:aII}) and (\ref{eq:solA}) for $\Phi=\calA_{\rm odd}$,
and using $M_S^2=(9/4)g_R^2W^2$ and parametrization (\ref{eq:param}), 
it is straightforward to find 
\begin{eqnarray}
L_{\rm sol}^3 &=& -\beta\Bigl(1+{\abs{\xi}^2\over V}\Bigr)
-{\alpha\over V}\Bigl({V-\abs{\xi}^2\over V+\abs{\xi}^2}\Bigr) \nn
V&=&{9\over2} g_R^2
g^{xy}{\partial W\over\partial\varphi^x}{\partial W\over\partial\varphi^y}
\left\{{1\over4}\Bigl[
\beta\Bigl(1+{\abs{\xi}^2\over V}\Bigr)-{\alpha\over V}\Bigr]^2+{\alpha\beta\over V}\right\} \nn
&&{}+{3\over2} g_R^2W^2
\left\{
{1\over2}\Bigl[\beta\Bigl(1+{\abs{\xi}^2\over V}\Bigr)-{\alpha\over V}\Bigr]^2
-{3\over2}\beta^2\Bigl(1+{\abs{\xi}^2\over V}\Bigr)
-{\alpha\beta\over V}\right\}.\hspace{2em}
\end{eqnarray}
We see that the brane tension $\mp3g_RWL^3_{\rm sol}$ with this 
$L_{\rm sol}^3$ coincides with Falkowski et al's\cite{ref:FLP} 
provided that 
$\xi$ terms are eliminated by assigning odd parity to $\xi$. 
Their tension parameter $\Lambda$ is identified with our $3g_RW$. 
If the system is reduced to the $n=0$ case where $W$ is constant, 
then this scalar potential $V$ also agrees with theirs aside from 
the term proportional to $\alpha\beta$, which we believe is their error or typo.

\section{Discussions}

We have given an off-shell formulation of the odd-parity 
`coupling constant' field $G(x)$ and 4-form multiplier field 
$H_{\mu\nu\rho\sigma}$. This was achieved by rewriting a neutral linear multiplet 
$\XL$ in terms of 3-form and 4-form gauge fields. 
In particular, we need a vector multiplet background in rewriting the 
auxiliary scalar component $N$ of $\XL$ into the 4-form gauge field 
$H_{\mu\nu\rho\sigma}$, and we obtain the 4-form field action in the 
five-dimensional bulk by applying the invariant action formula for 
the product of an Abelian vector multiplet $\XV$ and the linear multiplet 
$\XL$. Using here a vector multiplet different from the above background 
vector multiplet, we can obtain the `coupling constant' field $G(x)$ 
as the ratio of the two scalar components $M$ of the these two 
vector multiplets. All the components of this linear multiplet $\XL$ 
now become Lagrange multiplier fields and, in particular, the 
4-form gauge field component becomes the multiplier demanding that $G$ be 
a constant and change sign across the branes. 

We have presented this formulation in a rather general system of Yang-Mills 
 and hypermultiplet matters and discussed the relation between the bulk 
cosmological constant and the brane tensions of two boundary planes. 
The result agreed with the other authors' ones when the system is reduced
to some special cases. 

It is interesting that our approach suggests that this parity-odd 
coupling constant formulation cannot be generalized to the case of 
non-Abelian gauge coupling. This is because there appears no ratio of 
two scalar components which is group singlet to be identified with the 
coupling constant field if both of or one of the two vector multiplets 
are non-Abelian.

We have not discussed the other approach which Altendorfer et 
al\cite{ref:ABN} pursued. 
As discussed by Falkowski\cite{ref:FLP} and BKVP,\cite{ref:BKVP} 
this approach requires that the 
$U(1)_R$ generator $i\vec q\dt\vec\sigma$ should anti-commute with 
the matrix $M^i{}_j=(\sigma_3)^i{}_j$ which is used in defining $Z_2$ 
parity of the fermions. The off-shell formulation for this approach is 
not difficult and actually was given by Zucker\cite{ref:Zuk3} 
for the case of linear multiplet 
compensator in the bulk. Half components of the bulk compensator 
yield a compensator multiplet on the brane. Both linear multiplet and 
hypermultiplet compensator in the bulk induces a chiral multiplet 
compensator on the brane which we denote by $\Sigma_0$. Generally in 4D 
superconformal framework,\cite{ref:KU} 
the cosmological constant (without breaking supersymmetry) is supplied 
from the superpotential term $[\Sigma^3]_F$ of the chiral compensating 
multiplet $\Sigma$ with Weyl weight 1 provided that the $F$-component of
$\Sigma$ develops a non-zero VEV. The brane tension term here, 
therefore, should be produced by giving a superpotential $F$ term 
$[\Sigma_0^n]_F$ on the brane. ($n$ is a suitable power such that $n$ 
times Weyl weight of $\Sigma_0$ becomes 3.) But one can easily convince 
oneself that the $F$-component of this chiral compensator $\Sigma_0$ 
develops non-zero VEV if and only if the $U(1)_R$ generator $i\vec 
q\dt\vec\sigma$ anti-commutes with $\sigma_3$. Moreover, this 
superpotential term can be multiplied by any function $g(S)$ of (Weyl 
weight 0) matter chiral multiplets $S_i$ living on the brane as 
$[\Sigma_0^ng(S)]_F$, and then it will clearly yield the brane tension 
of arbitrary magnitude which has no relation with the bulk cosmological 
term.

\section*{Acknowledgements}
The authors would like to thank Hiroaki Nakano for his interest in this 
work. 
T.~K.\ is supported in
part by the Grants-in-Aid for Scientific Research No.~13640279 from 
Japan Society for the Promotion of Science and 
the Grants-in-Aid for Scientific Research on Priority Areas No.~12047214
from the Ministry of Education, Science, Sports and Culture, Japan.

\appendix

\section{Parametrization of hypermultiplet manifold}

We concentrate on only the odd $\alpha$ components $\calA^{\alpha=2\hat\alpha-1}_i$
since the even $\alpha$ components $\calA^{2\hat\alpha}_i$ are
given by the odd components by the reality as
$\calA^{2\hat\alpha}{}_i = (\calA^{2\hat\alpha-1}{}_j)^*\epsilon_{ji}$.
It was shown in I that the solution to the constraints 
(\ref{eq:YTconstr}) and (\ref{eq:EQmotion}) 
can be given in the form 

\begin{equation}
\calA_{\rm odd}\equiv\matrix{1 \cr 1\cr q \cr}\pmatrix{ 
\calA^{a=1}{}_{i=1} & \calA^{a=1}{}_{i=2} \cr
\calA^{a'=3}{}_{i=1} & \calA^{a'=3}{}_{i=2} \cr
\calA^{\ul\alpha={\rm odd}}{}_{i=1} & \calA^{\ul\alpha={\rm odd}}{}_{i=2} \cr}
= U_{2,q}
\pmatrix{ 
1 & 0 \cr
0 & 1 \cr
0 & 0 \cr}
\label{eq:solform}
\end{equation}
with a unitary matrix $U_{2,q}\in U(2,q)$, where 
$\calA^{\ul\alpha={\rm odd}}{}_i$ is a $q$-component complex vector 
for each $i=1,2$. This unitary matrix $U_{2,q}$ can be parametrized 
in the form\cite{ref:IKK}
\begin{eqnarray}
U_{2,q} &=& e^{\phi_i X^i}e^{\alpha^i(\phi,\phi^\dagger)X^\dagger_i}e^{\beta_m(\phi,\phi^\dagger)S_m}, \nn
\phi_i X^i&\equiv&
\pmatrix{ 
0 & 0 & 0 \cr
0 & 0 & 0 \cr
\phi_1 & \phi_2 & 0 \cr}, \qquad 
\alpha^i X^\dagger_i\equiv 
\pmatrix{ 
0 & 0 & \alpha^1 \cr
0 & 0 & \alpha^2 \cr
0 & 0 & 0 \cr}
\end{eqnarray}
by using two independent $q$-component vectors $\phi_1$ and $\phi_2$,
where $S_m$ are the generators of `unbroken' subgroup
$H=U(2)\times U(q)$ and $\beta_m$ are complex generally. 
(Note that the broken generators $X^i$,    
$X^\dagger_i$ and their coefficients $\phi_i$ and $\alpha^i$ are all 
$q$-components for each $i=1,2$.) 
Actually the first factor 
$e^{\phi_i X^i}$ parametrizes the complex group coset 
$G^C/\hat H$ ($G^C$: complex extension of $G$, $\hat H$: 
complex subgroup whose generators are given by $\{X^\dagger_i,\ S_m\}$), 
and is just the basic variable in the non-linear realization theory by 
BKMU\cite{ref:BKMU} in supersymmetric theory.
The coefficients $\alpha^i(\phi,\phi^\dagger)$ and $\beta_m(\phi,\phi^\dagger)$ are not independent 
parameters but are determined to be functions of $\phi_i$ and $\phi_i^\dagger$ 
by the demand that $U_{2,q}$ is a unitary matrix belonging to $U(2,q)$.  
It is not so easy to find explicit form of $\alpha^i(\phi,\phi^\dagger)$ and 
$\beta_m(\phi,\phi^\dagger)$, but we, fortunately here, can avoid the 
computation. Since $U_{2,q}$ is acting in Eq.~(\ref{eq:solform}) on the 
$(2+q)\times2$ matrix whose last $q$ raws are all zero and the first two
raws give $2\times2$ unit matrix, the third columns of the matrices 
$e^{\alpha^i(\phi,\phi^\dagger)X^\dagger_i}$ and 
$e^{\beta_m(\phi,\phi^\dagger)S_m}$ are irrelevant so that 
$e^{\alpha^i(\phi,\phi^\dagger)X^\dagger_i}$ can be replaced by 1 and 
$e^{\beta_m(\phi,\phi^\dagger)S_m}$ can be replaced by a $2\times2$ 
matrix $U^C$ acting from the right:
\begin{equation}
\calA_{\rm odd}=U_{2,q}
\pmatrix{ 
1 & 0 \cr
0 & 1 \cr
0 & 0 \cr}
= e^{\phi_i X^i}
\pmatrix{ 
1 & 0 \cr
0 & 1 \cr
0 & 0 \cr} U^C
= 
\pmatrix{ 
U^C \cr
(\phi_1\ \phi_2)U^C \cr}
\end{equation}
Now $U^C$ is determined by requiring that this form of $\calA_{\rm odd}$ 
satisfies the constraints (\ref{eq:YTconstr}) and (\ref{eq:EQmotion}), 
which is equivalent to the original condition that $U_{2,q}$ belongs 
to $U(2,q)$.
Then, imposing the $SU(2)$ and $U(1)_T$ gauge conditions on $U^C$ to 
take the form
\begin{equation}
U^C= 
\pmatrix{ 
a & 0 \cr
\calA^3{}_1 & b \cr}
\qquad (a,\,b: \hbox{real, positive})
\end{equation} 
we can find the solution of $U^C$  and hence of $\calA_{\rm odd}$ as given 
in Eq.~(\ref{eq:solA}) in the text.

The hypermultiplet scalar part of the Lagrangian is given as follows 
if written in the form before eliminating the auxiliary $SU(2)$ gauge 
field $V_\mu^{ij}$:
\begin{eqnarray}
e^{-1}{\cal L}_{\rm Hyp}&=&
+\calD^\mu\calA_i^{\bar \alpha}\calD_\mu\calA^i_\alpha 
=
-\abs{\calD_\mu\calA^a_i}^2
-|\calD_\mu\calA^{a'}_i|^2
+\abs{\calD_\mu\calA^{\ul\alpha}_i}^2
\label{eq:Hlag}
\end{eqnarray}
where 
\begin{eqnarray}
\calD_\mu\calA^a_i &=& \partial_\mu\calA^a_i+\calA^a_jV_\mu^{\,j}{}_i
-(W_{S\mu}+W_{T\mu})(i\sigma_3)^a{}_b\calA^b_i \nn
\calD_\mu\calA^{a'}_i &=& \partial_\mu\calA^{a'}_i+\calA^{a'}_jV_\mu^{\,j}{}_i
-W_{T\mu}(i\sigma_3)^{a'}{}_{b'}\calA^{b'}_i \nn
\calD_\mu\calA^{\ul\alpha}_i &=& \partial_\mu\calA^{\ul\alpha}_i
+\calA^{{\ul\alpha}}_jV_\mu^{\,j}{}_i
-W_{T\mu}(i\sigma_3)^{\ul\alpha}{}_{\ul\beta}\calA^{\ul\beta}_i 
-W_{\mu}^I(gt_I)^{\ul\alpha}{}_{\ul\beta}\calA^{\ul\beta}_i.
\end{eqnarray}
If we neglect the $U(1)_R$ gauge interaction and other matter 
gauge interaction with generators $gt_I$ by setting $g_R=g=0$, 
then the Lagrangian (\ref{eq:Hlag}) is simply rewritten in the following
form by using $(2+q)\times2$ complex matrix $\Phi=\calA_{\rm odd}$:
\begin{eqnarray}
&&e^{-1}{\cal L}_{\rm Hyp} =
2\tr[(D^\mu\Phi)^\dagger\eta(D_\mu\Phi)], \qquad 
D_\mu\Phi= \partial_\mu\Phi+\Phi A_\mu, \nn
&&(A_\mu)^i{}_j= (V_\mu)^i{}_j 
-iW_{T\mu}\delta^i_j ,\qquad 
\eta=\pmatrix{
-1 & 0 & 0 \cr
0 & -1 & 0 \cr
0 & 0 & {\bf1}_q \cr}
\label{eq:hyplag}
\end{eqnarray}
$A_\mu$ now gives a $U(2)$ gauge field which comes from the combination of 
the $SU(2)$ and $U(1)_T$ symmetries.
The four constraints (\ref{eq:YTconstr}) and (\ref{eq:EQmotion}) 
can be rewritten as 
\begin{equation}
\Phi^\dagger\eta\Phi= -{1\over2}{\bf1}_2.
\label{eq:constraint}
\end{equation}
The Lagrangian (\ref{eq:hyplag}) with constraints (\ref{eq:constraint}) 
clearly describes a nonlinear sigma model of Grassmannian manifold 
$U(2,q)/U(2)\times U(q)$.\cite{ref:BKY} 
If the $U(2)$ auxiliary gauge field $A_\mu$ is 
eliminated the lagrangian is rewritten as 
\begin{eqnarray}
&&e^{-1}{\cal L}_{\rm Hyp} =
2\tr[(\partial^\mu\Phi)^\dagger\eta(\partial_\mu\Phi)] +
4\tr[(\Phi^\dagger\eta\partial_\mu\Phi)^2].
\end{eqnarray}
If the expression (\ref{eq:solA}) 
of $\calA_{\rm odd}$ derived above is substituted for $\Phi$ here, 
then, it is easy to confirm that this Lagrangian can be 
written in the form
\begin{eqnarray}
e^{-1}{\cal L}_{\rm Hyp} &=&
{\partial K(\phi,\bar\phi)\over\partial\phi_I\partial\bar\phi_{\bar J}}
\partial_\mu\phi_I\partial^\mu\bar\phi_{\bar J}, \nn
K(\phi,\bar\phi) 
&=& -\ln\det\bigl[ 
\pmatrix{1 & 0 & \phi_1^\dagger\cr 
0 & 1 & \phi_2^\dagger\cr}\eta 
\pmatrix{1 & 0 \cr 
0 & 1 \cr 
\phi_1 & \phi_2 \cr} \bigr] \nn
&=& -\ln\bigl[ (1-\abs{\phi_1}^2)(1-\abs{\phi_2}^2)
-|{\phi_1^\dagger\cdot\phi_2}|^2 
\bigr]
\end{eqnarray}
with $\abs{\phi}^2\equiv\phi^\dagger\cdot\phi$. 
We see that this $K(\phi,\bar\phi)$ is just the well-known Zumino-Aoyama 
form\cite{ref:Zumino,ref:Aoyama} 
of the K\"ahler potential corresponding to the (K\"ahlerian) 
Grassmannian manifold $U(2,q)/U(2)\times U(q)$, giving a special 
example of the general form presented by BKMU.\cite{ref:BKMU}

If the $U(1)_R$ gauge field is retained in the calculation, we see that 
the derivative $\partial_\mu$ in this Lagrangian is replaced by the 
$U(1)_R$-covariant derivative $\nabla_{R\mu}=\partial_\mu-\delta_R(GW_{R\mu})$ 
with non-linear $U(1)_R$ transformation $\delta_R$. 

%

%

\end{document}